\documentclass[12pt]{article}
\usepackage{graphicx}
\usepackage{amssymb}
\usepackage{epstopdf}

\begin{document}

\title{Search for the $\theta^+(1540)$ in the Reaction $K^+p\to K^+n\pi^+$ at 11~GeV/$c$}
\author{{J.\ Napolitano, J.\ Cummings, and M.\ Witkowski$^\dagger$}\\
{\em Department of Physics, Rensselaer Polytechnic Institute, Troy, NY 12180}\\
{\small $^\dagger$ Current address: Goddard Space Flight Center, Greenbelt, MD}}

\maketitle

\begin{abstract}
We have studied the reaction $K^+p\to K^+n\pi^+$ using an 11~GeV/$c$ $K^+$ beam and the Large Acceptance Superconducting Solenoid (LASS) multiparticle spectrometer facility at SLAC. We put limits on the production of narrow $\theta^+$ baryons in this reaction.
\end{abstract}

There are numerous reports of experimental evidence for a narrow $S=+1$ baryon, the $\theta(1540)$, decaying to $K^+n$~\cite{PDGTheta,Hicks:2004vd,Nakano:2004cr,IUQNP}. Such an object is manifestly exotic; in terms of the quark model, it must be a $qqq\bar{s}q$ state.   These reports are not without controversy, however~\cite{PDGTheta,IUQNP,Dzierba:2003cm}, and there is much anecdotal evidence of searches that have yielded negative results but are left unreported except for conference talks. This paper describes such a negative result, originally presented at the 7th International Symposium on Meson - Nucleon Physics and the Structure of the Nucleon (MENU 97) Vancouver, Canada, 28 Jul - 1 Aug 1997~\cite{Napolitano:1997cd}.

\begin{figure}[t]
\begin{center}
\includegraphics[width=\textwidth]{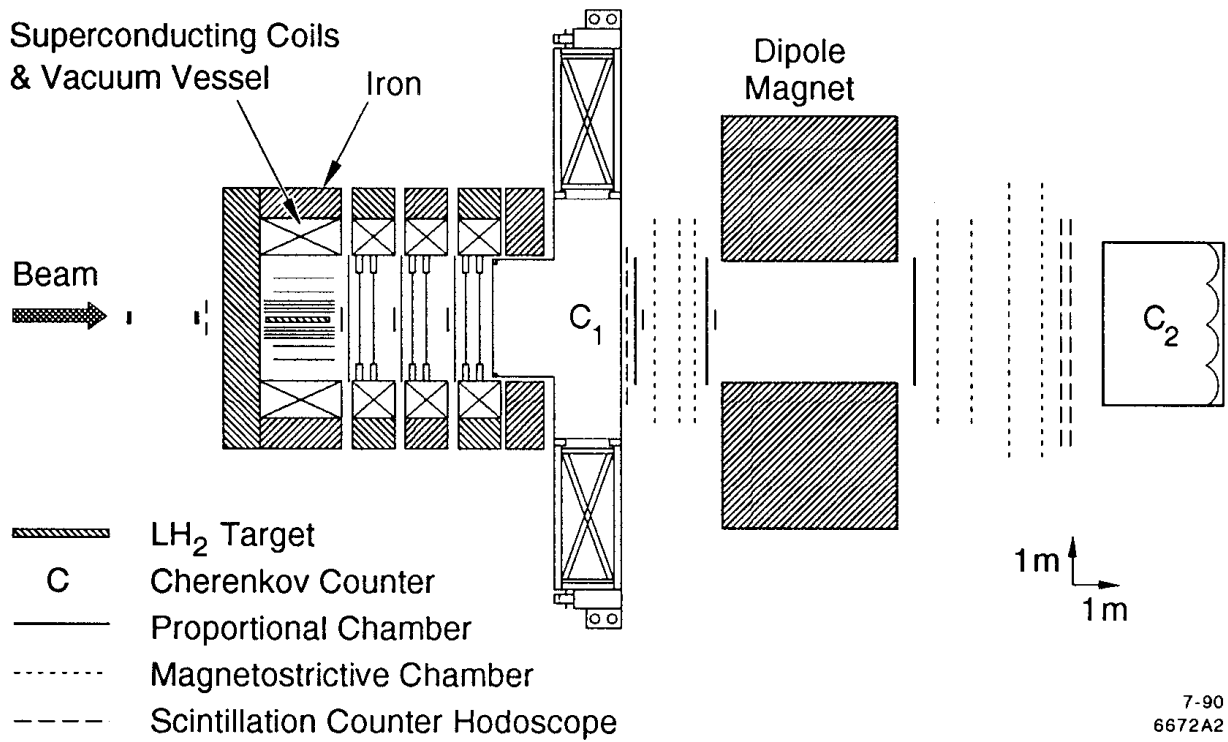}
\caption{The SLAC E135 LASS Spectrometer system.}
\label{fig:lass}
\end{center}
\end{figure}

This measurement used data from the Large Acceptance Superconducting Solenoid (LASS) spectrometer~\cite{lass} which provided very large and flat acceptance for mostly-charged final states. The SLAC E135 apparatus is shown in Fig.~\ref{fig:lass}. Its essential features are a 1.9~m diameter, 4.2~m long, 22.4~kG solenoidal magnetic field followed by a 30~kG$\cdot$m dipole; an 85~cm long liquid hydrogen target; six cylindrical PWC's, with anode and cathode readout, axially surrounding the target; six downstream planar PWC's which provide full-bore coverage; a set of magnetostrictive readout spark chambers for downstream tracking; scintillation counter hodoscopes; and two segmented Cherenkov counters filled with freon at atmospheric pressure. An RF-separated 11~GeV/$c$ $K^\pm$ beam impinged on a liquid hydrogen target after traversing a beam telescope which  yielded a precise measurement of individual beam particle momentum. Approximately 4~nb$^{-1}$ of $K^-$ data and 1~nb$^{-1}$ of $K^+$ data were acquired. A summary of the experimental results until 1990 is available~\cite{lassRes}.

The cylindrical PWC's made it possible to trigger on low multiplicity ($\leq2$) events with large transverse momentum in any number of tracks. This aspect of the experiment was crucial for many of the analyses, including the one described here.

The data were originally reduced to Data Summary Tapes (DST's) at SLAC and at Nagoya University.  The DST's contain all the information necessary to reconstruct the events according to a number of different topologies, and we used these for our analysis.  In fact, the DST's were originally further reduced at SLAC to a ``meson strip'' which required there be between two and nine reconstructed tracks, and that there be less than 2~GeV/c of missing longitudinal momentum.  The analysis described here began with these ``meson strip'' DST's.

We used the SLAC IBM mainframe to select events of interest, and to perform various constrained kinematic fits.  The results of these fits were stored as {\sc HBOOK} ntuples at SLAC, containing information on the fitted tracks, particle identification, and bookeeping data for each event. The ntuples were then transferred to Rensselaer for further analysis.

Our original motivation for searching for the $\theta(1540)$ (at the time, called a $Z$ baryon by the Particle Data Group) was motivated by the same work~\cite{Diakonov:1997mm} that has been the basis for the recent stimulation in this area. We inquired at that time, and the authors offered us an estimate of the production cross section in the reaction $K^+p\to\theta(1540)\pi^+$~\cite{polyPC} . This cross section would be about 10~${\mu}$b, and therefore well within reach of the LASS data set. We note that this reaction required identification of the final state $\pi^{+}_{\rm fast}K^{+}n$ where the pion is the forward particle recoiling against the (slowly moving) $K^{+}n$ system

Starting with the meson strip of the $K^+p$ DST's, we further required exactly two positively-charged tracks and a beam track to yield a production vertex within the target region.  The resulting data sample consisted mainly of $K^+p$ elastic scattering interactions. However, this background was readily removed using the obvious kinematic features in acoplanarity and the difference in the transverse momenta of the two tracks. Here, acoplanarity ${\cal A}$ is defined as the normalized dot product between the beam vector and the cross product of the vectors of the two outgoing particles.  That is,
$${\cal A}=\frac{\vec{p}_{\rm Beam}\cdot\left(\vec{p}_K\times\vec{p}_p\right)}
{\left|\vec{p}_{\rm Beam}\right|\left|\vec{p}_K\times\vec{p}_p\right|}$$
where $\vec{p}_K$ and $\vec{p}_p$ refer generically to the large angle and
small angle outgoing tracks, respectively. After removing elastic scattering events by means of a tight selection procedure which required that both the acoplanarity and the transverse momentum difference between the two tracks be near zero, the missing mass recoiling against $K^+\pi^+$ shows a clear neutron peak. The positively-charged particle with the larger momentum was then identified as a $\pi^+$ by means of a simple selection on the pulse height obtained from the downstream Cherenkov counter. Finally, interpreting the other charged track as a $K^+$, the missing mass recoiling against this $K^+$ and the leading $\pi^+$ was calculated, and required to be consistent with a missing neutron.

\begin{figure}
\begin{minipage}{3.0in}
\includegraphics[width=3.0in,height=3.0in]{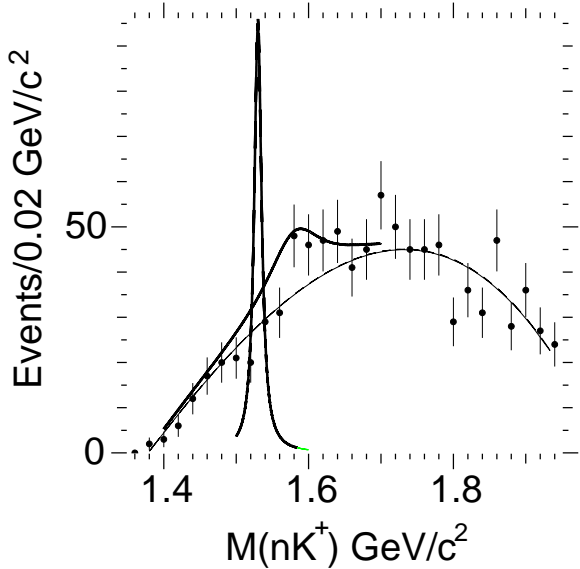}
\end{minipage}
\hfill
\begin{minipage}{3.0in}
\caption{Distribution of $K^+n$ mass in the reaction $K^+p\to\pi^+_{\rm fast}K^+n$.  The curves show the expected distributions based on the mass, width, and cross sections
for $\theta^+$ production and decay in this reaction.
See the text for details.}
\label{fig:Zlimits}
\end{minipage}
\end{figure}

The $K^{+}n$ effective mass spectrum is plotted in Fig.~\ref{fig:Zlimits}.
We have overlaid curves of $\theta^+$ production based on predictions~\cite{Diakonov:1997mm,polyPC} for a mass of 1540~MeV and a width of 12~MeV, and for a rudimentary fit allowing the mass and width to vary. Several assumptions are made in the preparation of this plot. First, we estimate the raw sensitivity of the LASS $K^{+}p$ data by scaling the published sensitivity of the $K^{-}p$ data by the ratio of the number of triggers. Second, we assume an overall efficiency of 10\%, which is clearly an underestimate based on simulations performed on other data sets. Also, in the absence of predictions, we assume a branching fraction of 50\% for $\theta^{+} \rightarrow K^{+}n$. The data shown in Fig.~\ref{fig:Zlimits} represent $\approx \frac{1}{3} $ of the total sample.

Using a production cross section of 10~${\mu}$b~\cite{polyPC}, our assumptions give us an expected signal of about 85 events.  This is used for both of the curves in Fig.~\ref{fig:Zlimits}.  While the broader solution cannot be ruled out, this analysis forbids the narrow solution. This is consistent with the absence of a signal in the $K_Sp$ mass distribution for the reaction $K^+p\to K_Sp\pi^+$~\cite{PDGTheta} at lower energies where the cross section is larger.

We gratefully acknowledge SLAC Group B, especially Bill Dunwoodie, Blair Ratcliff, and Dave Aston, for the use of their data and considerable assistance in the analysis.

\bibliographystyle{unsrt}

\begin{thebibliography}{99}

\bibitem{PDGTheta}
G.~Trilling, p.916, in S.~Eidelman {\it et al.}  [Particle Data Group Collaboration],
``Review of particle physics,''
Phys.\ Lett.\ B {\bf 592}, 1 (2004).
%%CITATION = PHLTA,B592,1;%%

\bibitem{Hicks:2004vd}
K.~Hicks,
``Experimental outlook for the pentaquark,'' arXiv:hep-ph/0408001 (2004).

\bibitem{Nakano:2004cr}
T.~Nakano and K.~Hicks,
``Discovery of the strangeness S = +1 pentaquark,''
Mod.\ Phys.\ Lett.\ A {\bf 19}, 645 (2004).

\bibitem{IUQNP}
See for example the 2004 International Conference on Quarks and Nuclear Physics, talks available at {\sf http://www.qnp2004.org/}.

\bibitem{Dzierba:2003cm}
A.~R.~Dzierba, D.~Krop, M.~Swat, S.~Teige and A.~P.~Szczepaniak,
``The evidence for a pentaquark signal and kinematic reflections,''
Phys.\ Rev.\ D {\bf 69}, 051901 (2004)

\bibitem{Napolitano:1997cd}
J.~Napolitano, J.~Cummings and M.~Witkowski, ``Baryon excitation in ${K}^\pm p$ reactions,'' $\pi{N}$ Newslett.\  {\bf 13}, 276 (1997).

\bibitem{lass} D.\ Aston, {\em et al.}, ``The LASS spectrometer'', SLAC-Rep-298, 1986.

\bibitem{lassRes} D.\ Aston, {\em et al.},
``Summary of LASS results and the future of strange quark spectroscopy", in Proceedings of the Rice Meeting of the Division of Particles and Fields of the APS, Vol.2, p.651 (1990).

\bibitem{Diakonov:1997mm}
D.~Diakonov, V.~Petrov and M.~V.~Polyakov,
``Exotic anti-decuplet of baryons: Prediction from chiral solitons,''
Z.\ Phys.\ A {\bf 359}, 305 (1997)

\bibitem{polyPC} M.\ Polyakov, private communication (1997).

\end{thebibliography}

\end{document}